\documentclass[english,prl,onecolumn,nofootinbib]{revtex4}
\usepackage{graphicx, subfig, bm, color, babel}
\usepackage{hyperref}
\usepackage{amsmath}
\usepackage{color}

\def\be{\begin{equation}}
\def\ee{\end{equation}}
\def\bea{\begin{eqnarray}}
\def\eea{\end{eqnarray}}

\def\d {\mathrm{d}}
\newcommand{\HH}{\mathcal{H}}
\newcommand{\D} {\nabla}

\renewcommand{\>}{\rangle}
\newcommand{\sfrac}[2]{{\textstyle\frac{#1}{#2}}}

\newcommand{\chiprime}{\tilde\chi}

\newcommand{\two}{^{\text{\tiny (2)}}}

\newcommand{\twod}{_{\text{\tiny (2)}}}

\newcommand{\p}{_{{\text{\tiny$\|$}}}}
\newcommand{\pp}{{{\text{\tiny$\|$}}}}
\renewcommand{\o}{{\hspace{-0.5mm}{\text{\tiny$\perp$}}}}
\renewcommand{\bot}{\o}

\newcommand{\Ic}{\int_0^{\chi_s}\!\!\!\d\chi\,}
\newcommand{\Icpr}{\int_0^{\chi}\!\!\!\d\chiprime\,}

\begin{document}

\title{Nonlinear relativistic corrections to cosmological distances,\\redshift and gravitational lensing magnification. I -- Key results}
\author{Obinna Umeh$^{1,2}$, Chris Clarkson$^1$ and  Roy Maartens$^{2,3}$\\
\emph{$^1$Astrophysics, Cosmology and Gravity Centre, and, Department of Mathematics and Applied Mathematics, University of Cape Town, Rondebosch 7701, South Africa.\\
$^2$Physics Department, University of the Western Cape,
Cape Town 7535, South Africa \\
$^3$Institute of Cosmology \& Gravitation, University of Portsmouth, Portsmouth PO1 3FX, United Kingdom\\
}}

\begin{abstract}

The next generation of telescopes will usher in an era of precision cosmology, capable of determining the cosmological model to beyond the percent level. For this to be effective, the theoretical model must be understood to at least the same level of precision. A range of subtle relativistic effects remain to be explored theoretically, and offer the potential for probing general relativity in this new regime. We present the distance-redshift relation to second order in cosmological perturbation theory for a general dark energy model. This relation determines the magnification of sources at high precision, as well as redshift space distortions in the mildly non-linear regime. We identify a range of new lensing effects, including: double-integrated and nonlinear integrated Sach-Wolfe contributions, transverse Doppler effects, lensing from the induced vector mode and gravitational wave backgrounds, in addition to lensing from the second-order potential. Modifications to Doppler lensing from redshift-space distortions are identified. Finally, we find a new double-coupling between the density fluctuations integrated along the line of sight, and gradients in the density fluctuations coupled to transverse velocities along the line of sight. These can be large and thus offer important new probes of gravitational lensing and general relativity. This paper accompanies \href{http://arxiv.org/abs/1402.1933}{Paper II}, where a comprehensive derivation is given. 

\end{abstract}

\maketitle

\paragraph{\bf\em Introduction}

The main probe of cosmological models comes from the relation between the angular diameter or luminosity distance of a source and its redshift. At its most basic level this relation determines the parameters of the cosmological model, but when perturbations due to structure are included, a host of new physics effects is revealed. The most important of these is the magnification of sources from over-densities, which arises as an integrated effect along our past lightcone. Another important property is the distortion in redshift space due to the radial motion of sources relative to the Hubble flow, which leads to a `Doppler lensing' phenomenon which has recently been explored~\cite{Bonvin:2008ni,Bolejko:2012uj,Bacon:2014uja}. More subtle are the integrated Sachs-Wolfe (ISW) terms due to the evolution of the potential along the line of sight, and the pure SW effect which arises from the potential difference between source and observer.

All of the known effects on the distance-redshift relation are calculated at linear order in perturbations~\cite{Bonvin:2005ps}. Recently, relativistic linear effects in the density contrast have been considered~\cite{Yoo:2010ni,Bonvin:2011bg,Challinor:2011bk,Bertacca:2014dra}, and also in redshift space distortions~\cite{Bertacca:2012tp}. However, at second order other general relativistic effects must come into play. As structure evolves, the linear modes generate nonlinear ones, many of which are not present in Newtonian theory. Important dynamical examples of these are the induced vector and tensor modes whose spectra peak in power at the equality scale~\cite{Lu:2008ju,Baumann:2007zm}. The full effect of such modes on the optical properties of the model are not known in detail, though there are partial second-order results which include the distance-redshift relation Taylor expanded to $\mathcal{O}(z^2)$ in a pure dust model~\cite{Barausse:2005nf}, and in LCDM~\cite{Clarkson:2011uk}. More generally, the lensing shear is given to second-order in~\cite{Bernardeau:2009bm,Bernardeau:2011tc}. We present here the second-order distance-redshift relation for a general dark energy model, and identify key new lensing effects, some of which will be observable with the next generation of cosmological experiments. The derivation is very involved and is given in an accompanying paper~\cite{Umeh:2014ana}.

Another group independently presented the full second-order distance-redshift relation using a very different approach~\cite{BenDayan:2012wi}~-- see also related work~\cite{BenDayan:2012ct,BenDayan:2013gc,Fanizza:2013doa,Nugier:2013tca}. Their approach starts with a general light-cone coordinate system, from which the relevant relations are found using a suitable coordinate transformation. Our approach solves the Sachs equations perturbatively in the familiar Poisson (Newtonian) gauge. Our final result is presented in a fully expanded, ready-to-use and  familiar form~-- as Doppler terms at the source, and as integrals over the Newtonian potential, with terms grouped into separate physical effects.

\paragraph{\bf\em Nonlinear optical equations}

In a general spacetime, the optical properties are governed by the Sachs equations. Lensing effects comes in several parts. The redshift of a source $z$ is related to the affine parameter $\lambda$ along a light ray via~\cite{Clarkson:2011br}
\be\label{energyprorpa1}
\frac{\d  z}{\d \lambda}=-(1+ z)^2\left[\frac{1}{3}\, \Theta + \sigma_{ab}n^{a}n^{b}\right],
\ee
where $\Theta$ is the volume expansion rate of cold matter, $\sigma_{ab}$ is its shear, and $n^a$ is the spatial direction of the light ray which we choose pointing in the direction of observation.
Lensing magnification and shear are determined by
\begin{eqnarray}\label{Areadistance}
\frac{\d^2 D_A}{\d\lambda^2}&=&-\left[\frac{1}{2}  R_{ab} k^a k^b+ {\Sigma}_{ab}\Sigma^{ab}\right] D_A\,,~~~
\frac{\d{\Sigma}_{\<ab\>}}{\d\lambda}=-{\Sigma}_{ab}{\theta}+ N_{\<e}{}^{a} N_{f\>}{}^{c}{R}_{abcd} k^c k^d.
\end{eqnarray}
Here, $D_A$ is the area or angular diameter distance, $k^a$ is a future-pointing null vector with expansion $\theta$ and shear $\Sigma_{ab}$, $N_{ab}$ is the screen space metric, 
and angle brackets denote the trace-free part in the screen space. We expand and integrate these equations with the appropriate boundary conditions to second order in perturbation theory to determine redshift perturbations and lensing magnification to high precision in a perturbed flat FLRW model, without assumptions on the matter content.

\paragraph{\bf\em Second order perturbations}

For a flat background the metric in the Poisson gauge is\footnote{In our accompanying `derivations' paper we work on a perturbed Minkowski geometry conformally related to the FLRW one which we denote with a hat~-- we do not refer to the Minkowski background here, so remove the hats to avoid clutter. We have defined the redshift and distance perturbations below with respect to the Minkowski background.}
\begin{eqnarray}\label{metric}
\d s^2 &=&a^2\left[-(1 + 2\Phi+\Phi\two )\d \eta^2 + 2\omega_{i} \d x^{i}\d \eta
 + \left((1-2 \Phi-\Psi\two)\delta_{i j} + h_{ij}\right)\d x^{i}\d x^{j}\right],
\end{eqnarray}
where 
we neglect first-order anisotropic stress and vector and tensor modes. The observer 4-velocity $u^a$ is
\bea
u^0 &=&\frac{1}{a}\left( 1 - \Phi -\frac{1}{2}\Phi\two + \frac{3}{2}\Phi^2 + \frac{1}{2}\nabla_iv\nabla^iv \right),
~~~~
u^i=\frac{1}{a}\left( \nabla^iv+\frac{1}{2}\nabla^iv\two+\frac{1}{2}v\twod^i\right)
\eea
where $v$ is the first-order velocity potential. The second-order velocity has scalar $v\two$ and vector $v\two_i$ contributions.

\paragraph{\bf\em Perturbative calculation overview}
 The redshift perturbation is given by
\begin{equation}\label{eq:forredshiftexp}
(1+{z}_s) = (1+\bar{z})\left( 1 +\delta z +\sfrac{1}{2}\delta^2 z\right)\,,
\end{equation}
where $z_s$ denotes the observed position of the source, $\bar z$ is the background redshift and $\delta^n z$ is the $n$-th order redshift perturbation evaluated at the background position of the source $\lambda_s$.  These are found by perturbatively solving \eqref{energyprorpa1} as a function of the background affine parameter $\lambda$. 
Then, the distance is found from expanding \eqref{Areadistance}
\begin{eqnarray}\label{Areadistance2}
  D_A(\lambda_s)&=& a(\lambda_s)\left[(\lambda_o-\lambda_s)+ {\delta D_A}(\lambda_s)+ \sfrac{1}{2}{\delta^2 D_A}(\lambda_s)\right]\,.
\end{eqnarray}
Once the distance perturbations $\delta^nD_A$ are found, \eqref{eq:forredshiftexp} and~\eqref{Areadistance2} form a parametric relationship for $D_A(z_s)$~-- this is the form suitable for raytracing through specific objects in an N-body simulation, for example. To give $D_A(z_s)$ explicitly, we eliminate $\lambda_s$ perturbatively 
in favour of a new affine parameter $\chi_s$ in the background along the null ray, which is also the co-moving distance to the source calculated for an observed redshift $z_s$ using the background relationship
\be
{\chi_s} 
 = \int_{0}^{z_s} \frac{\d z}{(1+z)\HH(z)}\,,
 \ee 
 and $\HH=a'/a$ is the conformal Hubble rate. We then find
   \begin{eqnarray}\label{DAgen2}
 D_A(z_s)&=&\frac{\chi_s}{1+z_s} \left\{1+\left[\frac{\delta D_A}{\chi_s}+\left(1- \frac{1}{\HH_s \chi_s}\right)\delta z\right]
 +
 \frac{1}{2}\left[\frac{ \delta^2 D_A}{\chi_s} +\left(1- \frac{1}{\HH_s \chi_s}\right)\delta^2 z
  \right. \right.\\ \nonumber&&\left.\left.
 + 2\left(\frac{\delta D_A}{\chi_s}-\frac{1}{\HH_s}\frac{\d\delta z}{\d\chi}\right) \left(1- \frac{1}{\HH_s \chi_s}\right)\delta z-2\frac{\delta z}{\HH_s}\frac{\d}{\d\chi}\frac{\delta D_A}{\chi} 
 +  \left(
 \frac{\HH'_s}{\HH^2}- 1\right)\frac{(\delta z)^2}{\HH_s\chi_s}\right]
 \right\}.
 \end{eqnarray}
This  is  the suitable expression for further analysis on spheres of constant redshift.

\paragraph{\bf\em Redshift}

The first-order contribution to the redshift is
\be
{\delta z}
=\D\p v_s-\D\p v_o-\Phi_s+\Phi_o -2\Ic  \Phi'\,.
\ee
Here, $\chi$ is the affine parameter 
in the background along the null ray, and is also the co-moving distance to the source calculated for an observed redshift $z_s$.
The radial derivative along the null ray is $\nabla{\p}=n^i\nabla_i=\partial_\eta+\d/\d\chi$, where $n^i$ is the direction to the source in the background, and $\d/\d\chi$ is the derivative down the past lightcone. 
The redshift perturbation
consists of two boundary terms in the form of a Doppler term and a Sachs-Wolfe (SW) contribution from the difference in potentials at source and observer, as well as an integrated Sachs-Wolfe (ISW) term. At second order we have similar contributions:
\be
{\delta^2 z}=\mbox{Doppler}\two+\mathrm{SW}+ \mathrm{Doppler}\times\mathrm{SW}+\mathrm{ISW}\,.
\ee
The pure Doppler terms consist of pure second-order contributions, and quadratic first-order terms, which include the transverse Doppler effect, and a new coupling between source and observer:
\bea
&&\hspace{-4mm}\mathrm{Doppler}\two=v_{\|s}\two -v_{\|o}\two +\D{\p} v_s\two - \D{\p} v_o\two+\left(\D\p v_o-\D\p v_s\right)^2+\nabla_{\bot k}v_s\nabla_{\bot }^k v_s-\nabla_{\bot k}v_o\nabla_{\bot }^k v_o
\eea
Here, we have defined $v_{\|}\two=n^iv_{i}\two$ as the radial part of the vector mode of the velocity and $\nabla_{\bot i}=N_i^j\D_j$ as the covariant derivative on the screen space.
The SW terms also contain pure and mixed contributions, which couple the potential at the source and observer:
\be
\mathrm{SW}= -\Phi_s\two+\Phi_o\two +(\Phi_s-\Phi_o)\left(3\Phi_s+\Phi_o\right)\,.
\ee
There is a coupling between the Doppler terms at source and observer with the potential:
\bea
\mathrm{Doppler}\times\mathrm{SW}&=&-4\left(\Phi_s\D\p  v_s- \Phi_o  \D_{\| }v_o\right)
+2 \left( \Phi_s\D\p  v_o- \Phi_o\D\p v_s\right)\,.
\eea
The ISW effect at second order is much more complicated, and consists of several contributions:
\bea
\mathrm{ISW}&=&\text{ ISW}^{(2)} +\text{ISW}\times\text{SW}
+\text{ISW}\times\text{Doppler}
+\text{Integrated ISW}
\eea
where we have a pure second-order ISW contribution:
\be
\text{ ISW}^{(2)}=
-\Ic\left( {\Phi\two}' + {\Psi\two}' 
+{\omega\p\two}'- {h\p\two}'-8\Phi\Phi'\right)\, .
\ee
We have defined $h_\|=n^in^jh_{ij}$ as the radial part of the tensor mode.
Then we have the first-order ISW effect coupled with SW and Doppler terms:
\begin{eqnarray}
&&\text{ISW}\times\text{SW}= 4\left(\Phi_s-\Phi_o\right) \Ic\Phi'\,,\\
&&\text{ISW}\times\text{Doppler}=-4\left(\D\p v_s-\D\p v_o\right) \Ic\Phi'-4 \nabla_{\bot i} v_s \Ic \nabla_{\bot}^i \Phi \,.
 \end{eqnarray}
Finally, we have the double- and tripple-integrated SW terms:
\begin{eqnarray}
&&\text{Integrated ISW}= +8\Ic\Phi'\Icpr\Phi'(\chiprime)-8\Ic
\Icpr\frac{\chiprime}{\chi}\nabla_{\o}^i\Phi'(\chiprime)\Icpr\nabla_{\o i}\Phi(\chiprime)\,.
 \end{eqnarray}

These new effects contribute to the nonlinear redshift space distortions.

\paragraph{\bf\em Distance-redshift relation}

We define the fractional fluctuation in the area distance at observed redshift $z_s$ as 
\be
\Delta(z_s)=\frac{D_A(z_s)-\bar D_A(z_s)}{\bar D_A(z_s)}
\ee
where the background area distance is
$
 \bar{D}_A(z_s) 
 ={\chi_s}/{(1+{z}_s)}.
$
 \  The perturbation $\Delta$ splits into local terms evaluated at either the source or observer, local times integrated terms, and integrated terms:
 \be
 \Delta=\Delta_\text{loc}+\Delta_\text{loc-int}+\Delta_\text{int}\,.
 \ee
 In order to consistently make this split we have systematically replaced all radial derivatives which appear in integrated terms with partial time and proper null derivatives, and performed integration by parts to eliminate all null derivatives. 
\\ 
 
\noindent{\sc local terms:} These have Sachs Wolfe and Doppler contributions, with cross-terms between the two:
\be
\Delta_\text{loc}=\Delta_\text{loc}^\Phi+\Delta_\text{loc}^v+\Delta_\text{loc}^{\Phi\times v}\,.
\ee
Each of these further splits into terms which are purely local to the observer, or to the source, or coupling between the two. For the Sachs-Wolfe terms we have
\bea
\Delta_\text{loc,o}^\Phi&=&\left(1- \frac{1}{\HH_s \chi_s}\right)\Phi_o+\frac{1}{2}\left(1- \frac{1}{\HH_s \chi_s}\right)\Phi\two_o-\frac{1}{2}\left(1- \frac{1}{\HH_s \chi_s}\right)\omega_{\p o}-\frac{1}{2}\left(\frac{\HH'}{\HH^2}-7\right)\Phi_o^2 \,,\\
\Delta_\text{loc,s}^\Phi&=&-\left(2- \frac{1}{\HH_s \chi_s}\right)\Phi_s-\frac{1}{2} \Psi\two_s-\frac{1}{2}\left(1- \frac{1}{\HH_s \chi_s}\right)\Phi_s\two+\frac{1}{4}\left(1- \frac{2}{\HH_s \chi_s}\right)\omega_{\pp s}-\frac{1}{4} h_{\pp s}
\\ \nonumber&&
-\frac{1}{2}\left(\frac{\HH'}{\HH^2}+\frac{2}{\chi_s\HH_s}-7\right)\Phi_s^2
-\left(1-\frac{2}{\chi_s\HH_s}\right)\chi_s\Phi_s\Phi'_s-\chi_s\Phi_s\D_{\pp}\Phi_s
 \,,\\
\Delta_\text{loc,o-s}^\Phi&=&\left[-\left(\frac{\HH'}{\HH^2}-\frac{1}{\chi_s\HH_s}+4\right)\Phi_s+\left(1-\frac{2}{\chi_s\HH_s}\right)\chi_s\Phi'_s-\chi_s\D_{\pp}\Phi_s\right]\Phi_o\,.
\eea
The Doppler terms are
\bea
\Delta_\text{loc,o}^v&=&\frac{1}{\HH_s \chi_s}\D_{\pp}v_o+\frac{1}{2\HH_s \chi_s}\D\p  v_o\two-\frac{1}{2\HH_s\chi_s}v_{\p o}\two-\left(\frac{\HH'}{2\HH^2}-1\right)\D_{\pp}v_o\D_{\pp}v_o  + \frac{1}{\chi_s\HH_s}\nabla_{\o i}v_o\nabla_{\o}^iv_o\,,\\
\Delta_\text{loc,s}^v&=&\left(1- \frac{1}{\HH_s \chi_s}\right)\D_{\pp}v_s+\frac{1}{2}\left(1- \frac{1}{\HH_s \chi_s}\right)\D\p  v\two_s+\frac{1}{2}\left(1- \frac{1}{\HH_s \chi_s}\right)v_{\p s}\two
+\frac{1}{2}\left(\frac{\HH'}{\HH^2}-\frac{2}{\chi_s\HH_s}-1\right)\D_{\pp}v_s\D_{\pp}v_s
\nonumber\\&& +\chi_s\left(1-\frac{1}{\chi_s\HH_s}\right)\D_{\pp} v_s
\left(\D_{\pp}v'_s-\D_{\pp}^2v_s \right)+\left(1-\frac{1}{\chi_s\HH_s}\right)\nabla_{\o i}v_s\nabla_{\o}^iv_s\,,\\
\Delta_\text{loc,o-s}^v&=&\left[-\left(\frac{\HH'}{\HH^2}-\frac{1}{\chi_s\HH_s}\right)\D_{\pp}v_s-\chi_s\left(1-\frac{1}{\chi_s\HH_s}\right)\left(\D_{\pp}v'_s
-\D_{\pp}^2v_s \right)\right]\D_{\pp} v_o\,.
\eea
These contain a mixture of radial and transverse velocity terms, as well as the redshift space distortion term proportional to $\left(\D_{\pp}v'_s
-\D_{\pp}^2v_s \right)$. The mixed terms are 
\bea
\Delta_\text{loc,o}^{\Phi\times v}&=&-\left(\frac{\HH'}{\HH^2}+\frac{1}{2\chi_s\HH_s}+\frac{11}{2}\right)\Phi_o\D_{\pp}v_o\,, \\
\Delta_\text{loc,s}^{\Phi\times v}&=&-\left(\frac{\HH'}{\HH^2}-\frac{1}{\chi_s\HH_s}+3\right)\Phi_s\D_{\pp}v_s+\left(\frac{\HH'}{\HH^2}-1\right)\Phi_o\D_{\pp}v_s+\chi_s\left(1-\frac{2}{\chi_s\HH_s}\right)\Phi'_s\D_{\pp} v_s
\\ \nonumber&&
+\chi_s\D_{\pp}\Phi_s\D_{\pp}v_s-\chi_s\left(1-\frac{1}{\chi_s\HH_s}\right)\Phi_s(\D_{\pp}v'_s-\D_{\pp}^2v_s)\,,\\
\Delta_\text{loc,o-s}^{\Phi\times v}&=&\left[-\chi_s\left(1-\frac{2}{\chi_s\HH_s}\right)\Phi'_s
 -\chi_s\D_{\pp}\Phi_s
+\left(\frac{\HH'}{\HH^2}+2\right)\Phi_s\right]\D_{\pp}v_o+\left(1-\frac{1}{\chi_s\HH_s}\right)\chi_s\Phi_o(\D_{\pp}v'_s-\D_{\pp}^2v_s) \,.
\eea 
\\
 
 \noindent{\sc local-integrated terms:} 
Integrated terms come coupled to the local potential and Doppler terms
\be
\Delta_\text{loc-int}=\Delta_\text{loc-int}^\Phi+\Delta_\text{loc-int}^v,
\ee
which in turn have contributions from both observer and source. First the terms coupled to the potential:
\bea
\Delta_\text{loc-int,o}^\Phi&=&-\Phi_o\left[\left(1+\frac{2}{\chi_s\HH_s}\right)\frac{2}{\chi_s}\Ic\Phi+2\left(\frac{\HH'}{\HH^2}+\frac{3}{\chi_s\HH_s}-4\right)\Ic\Phi'
-\left(2+\frac{1}{\chi_s\HH_s}\right)\Ic\chi\nabla_{\o}^2\Phi
\right.\\ \nonumber&&\left.
+\left(9+\frac{2}{\chi_s\HH_s}\right)\Ic\frac{(\chi-\chi_s)\chi}{\chi_s}\nabla^2_{\o}\Phi
+ 4\Ic\frac{\chi}{\chi_s}\Phi'
  +4\Ic\frac{(\chi-\chi_s)}{\chi_s}
  \Phi'-4\Ic\frac{(\chi-\chi_s)\chi}{\chi_s}\Phi'' 
    \right]\,,\\
\Delta_\text{loc-int,s}^\Phi&=& \Phi_s\bigg[\left(1-\frac{1}{\chi_s\HH_s}\right)\frac{4}{\chi}\Ic\Phi+2\left(\frac{\HH'}{\HH^2}-\frac{2}{\chi_s\HH_s}+2\right)\Ic\Phi'
-\left(2+\frac{1}{\chi_s\HH_s}\right)\Ic\chi\nabla_{\o}^2\Phi
\\ \nonumber&&
+2\left(1-\frac{1}{\chi_s\HH_s}\right)\Ic\frac{(\chi-\chi_s)\chi}{\chi_s}\nabla^2_{\o}\Phi\bigg]-2\bigg[\left(1-\frac{2}{\chi_s\HH_s}\right)\chi_s^2\Phi'_s
+\chi_s\D_{\pp}\Phi_s\bigg]\Ic\Phi'\,.   
\eea 
These contain terms with the potential at observer and source coupled to the usual gravitational lensing  potential, as well as other integrated terms which appear at first-order. For the Doppler terms we have 
\bea
\Delta_\text{loc-int,o}^v&=&\D_{\pp}v_o\bigg[\frac{2}{\chi_s}\left(1-\frac{2}{\chi_s\HH_s}\right)\Ic\Phi-2\Ic\Phi'
+\left(2-\frac{1}{\chi_s\HH_s}\right)\Ic\frac{(2\chi-3\chi_s)\chi}{\chi_s}\nabla^2_{\o}\Phi
\bigg]\,,\\
\Delta_\text{loc-int,s}^v&=&-\D_{\pp}v_s
\bigg[\frac{2}{\chi_s}\left(5-\frac{1}{\chi_s\HH_s}\right)\Ic\Phi-4\left(1-\frac{1}{\chi_s\HH_s}\right)\Ic\Phi'
\\ \nonumber&&
-\left(2-\frac{1}{\chi_s\HH_s}\right)\Ic\chi\nabla_{\o}^2\Phi
+\left(3-\frac{2}{\chi_s\HH_s}\right)\Ic\frac{(\chi-\chi_s)\chi}{\chi_s}\nabla^2_{\o}\Phi
\bigg]
\\ \nonumber&&
+2\left(1-\frac{1}{\chi_s\HH_s}\right)\nabla_{\o} v_s\Ic\nabla_{\o}^i\Phi
-2\left(1-\frac{1}{\chi_s\HH_s}\right)\left[\chi_s\left(\D_{\pp}v'_s-\D_{\pp}^2v_s\right)\right]
\Ic\Phi'\,.
\eea
The second-last term is a coupling between the transverse velocity and the deflection angle to the source, while the final term is a coupling between the linear ISW effect and redshift space distortions. \\

 \noindent{\sc integrated terms:} 
A variety of new effects appear in the form of single, double or triple-integrated terms. We group these as direct integrated terms~-- including effects from time delay, first- and second-order ISW terms, and first- and second-order standard lensing convergence terms~-- then  double and triple integrated terms, which include integrated couplings of bending angle with convergence and ISW terms (there are no integrated Doppler contributions): 
\be
\Delta_\text{int}=\Delta_\text{int}^\text{direct}+\Delta_\text{int}^\text{multiple}\,.
\ee
The direct integrated terms are just extensions of the first-order distance redshift modifications to second-order:
\begin{eqnarray}
\Delta_\text{int}^\text{direct}&=& \frac{2}{\chi_s}  \Ic \Phi+\frac{1}{2\chi_s} \Ic(\Phi\two+ \Psi\two)-\frac{1}{2\chi_s} \Ic\frac{(\chi-\chi_s)}{\chi} \omega_{\p} 
+ \frac{3}{2\chi_s}\Ic \frac{(\chi-\chi_s)}{\chi} h_{\p} 
 \\
\nonumber&&-2\left(1- \frac{1}{\HH_s \chi_s}\right)\Ic  \Phi'
-\frac{1}{2}\left(1- \frac{1}{\HH_s \chi_s}\right)\Ic\left( \Phi\twod' + \Psi\twod'\right)
\\ \nonumber&&-\frac{1}{2}\Ic\left(1- \frac{1}{\HH_s \chi_s}+\frac{\left(2\chi-\chi_s\right)}{2\chi_s}\right)\omega\p'
+\frac{1}{2}\Ic \left(1- \frac{1}{\HH_s \chi_s}-\frac{\left(2\chi-\chi_s \right)}{\chi_s}\right)h\p'\\
\nonumber&&
+ \Ic \frac{{(\chi-\chi_s) \chi}}{\chi_s}\nabla^2_{\bot}\Phi +\frac{1}{4}\Ic \frac{(\chi-\chi_s)\chi}{\chi_s} \nabla_{\bot}^2( \Phi\two+\Psi\two)-\frac{1}{4}\Ic\frac{ (\chi-\chi_s)\chi}{\chi_s}\nabla_{\bot}^2\omega_{\p}\\\nonumber&&
- \frac{1}{4} \Ic\frac{(\chi-\chi_s)\chi}{\chi_s} \nabla_{\bot}^2 h_{\p}\,.
\end{eqnarray}
The first line contains time-delay contributions, the second and third are the `pure' ISW terms from scalars, vectors and tensors, while the final two lines contain the usual lensing terms, which are integrals over the screen-space Laplacian of all the radial metric potentials. (These are essentially all extensions of the first-order expression to second-order.) 
The multiple integrated non-linear terms are our key new results, and we give them according to whether they involve the first-order lensing convergence $\kappa$, bending angle $\alpha$ or shear $\gamma$: 
\be
\Delta_\text{int}^\text{multiple}=\Delta_\text{int}^{\Phi-\Phi}+\Delta_\text{int}^{\Phi-\kappa}+\Delta_\text{int}^{\alpha-\alpha}+\Delta_\text{int}^{\kappa-\kappa}+\Delta_\text{int}^{\alpha-\nabla\kappa}+\Delta_\text{int}^{\Sigma-\Sigma}\,.
\ee
Integrals over the potential and its time derivative couple:
\begin{eqnarray}
\Delta_\text{int}^{\Phi-\Phi}&=&  -\Ic\frac{3\chi-2\chi_s}{\chi_s}\Phi^2+4\Ic\frac{(2\chi-\chi_s)}{\chi_s}\Phi\Phi'
  \\ \nonumber &&
 + 2\left[\frac{2}{\chi_s}\left(3-\frac{2}{\chi_s\HH_s}\right)\Ic\Phi+\left(\frac{\HH'}{\HH^2}+\frac{1}{\chi_s\HH_s}-2\right)\Ic\Phi'
\right]\Ic\Phi'
\\ \nonumber&&
 +4\Ic\frac{(\chi-\chi_s)}{\chi\chi_s}\Phi'\Icpr\Phi(\chiprime)  -2\Ic\frac{2\chi-\chi_s}{\chi^2\chi_s}\Phi\Icpr\Phi(\chiprime)
  +4\Ic\Phi'\Icpr\Phi'(\chiprime)
\,.
\eea
First-order lensing-type contributions (i.e., integrals over $\nabla_\o^2\Phi$) couple to integrals over the potential and ISW terms:
\bea
\Delta_\text{int}^{\Phi-\kappa}&=&  -2\Ic\frac{1}{\chi}\Phi\Icpr\frac{(\chiprime-\chi)\chiprime}{\chi}\nabla_{\o}^2\Phi(\chiprime)
  -2\Ic\frac{(\chi-\chi_s)}{\chi^2\chi_s}\Phi\Icpr{(\chiprime-\chi)\chiprime}\nabla_{\o}^2\Phi(\chiprime)
    \\ \nonumber&&
  +\Ic\frac{(9\chi-2\chi_s)(\chi-\chi_s)}{\chi_s}\Phi\nabla^2_{\o}\Phi
  -2\Ic\Phi\Icpr\frac{\chiprime}{\chi}\nabla_{\o}^2\Phi(\chiprime)
  +2\Ic\frac{(\chi-\chi_s)}{\chi_s}\nabla_{\o}^2\Phi\Icpr\Phi(\chiprime)
  \\ \nonumber&&  
  +2\Ic\frac{(\chi-\chi_s)}{\chi_s}\Phi'\Icpr\chiprime\nabla_{\o}^2\Phi(\chiprime)
  +2\Ic\frac{(\chi-\chi_s)}{\chi}\Phi\Icpr\frac{(\chiprime-\chi)\chiprime}{\chi_s}\nabla_{\o}^2\Phi(\chiprime)\\ \nonumber&&
 + 2\left[
-\left(2-\frac{1}{\chi_s\HH_s}\right)\Ic\chi\nabla^2_{\o}\Phi
+\left(3-\frac{2}{\chi_s\HH_s}\right)\Ic\frac{(\chi-\chi_s)\chi}{\chi_s}\nabla^2_{\o}\Phi\right]\Ic\Phi'
  \,.
  \eea
Next, terms related to the bending angle from the potential couple together:   
  \bea
\Delta_\text{int}^{\alpha-\alpha}&=&  -4\Ic\frac{(\chi-\chi_s)}{\chi_s}\nabla_{\o i}\Phi\Icpr\frac{\chiprime}{\chi}\nabla_{\o}^i\Phi(\chiprime)
  -4\Ic\frac{\chi}{\chi_s}\nabla_{\o i}\Phi\Ic\nabla^i_{\o}\Phi
\\ \nonumber&&  
  -\frac{4}{\chi_s}\Ic\Icpr\nabla_{\o i}\Phi(\chiprime)\Icpr\nabla_{\o}^i\Phi(\chiprime)
  +6\Ic\frac{(\chi-\chi_s)\chi}{\chi_s}\nabla_{\o i}\Phi\nabla_{\o}^i\Phi
  \\ \nonumber&&
   -4\Ic\frac{(\chi-\chi_s)\chi}{\chi_s}\nabla_{\o i}\Phi'\Icpr\nabla_{\o}^i\Phi(\chiprime)
  -4\left(1-\frac{1}{\chi_s\HH_s}\right)
\Ic\frac{\chi}{\chi_s}\nabla_{\o}^i\Phi'\Icpr\nabla_{\o}^i\Phi(\chiprime)\,.
\eea
Lensing-lensing terms couple:
\bea\label{sdkflbjvskdvbs}
\Delta_\text{int}^{\kappa-\kappa}&=&  -\Ic\frac{(\chi-\chi_s)}{\chi_s}\nabla_{\o}^2\Phi\Icpr(\chiprime-\chi)\chiprime\nabla_{\o}^2\Phi(\chiprime)\,.
\eea
Then, a closely related contribution is from bending angle coupled to angular gradients of the lensing contribution:
\bea
\Delta_\text{int}^{\alpha-\nabla\kappa}&=&   +4\Ic\frac{(\chi-\chi_s)}{\chi_s}\Icpr\nabla_{\o i}\Phi(\chiprime)\Icpr\frac{{\chiprime}^2}{\chi}\nabla_{\o}^i\nabla_\o^2\Phi(\chiprime) 
\\ \nonumber&& 
-2\Ic\frac{(\chi-\chi_s)}{\chi_s}\nabla_{\o i}\Phi\Icpr\frac{(\chiprime-\chi){\chiprime}^2}{\chi}\nabla_{\o}^i\nabla_{\o}^2\Phi(\chiprime)\,.
\eea
Finally we have the contributions from the shear of the null geodesics in the screen space:
\bea
\Delta_\text{int}^{\Sigma-\Sigma}&=& -2\Ic\frac{(\chi-\chi_s)\chi}{\chi_s}\Icpr\nabla_{\o\<i}\nabla_{\o j\>}\Phi(\chiprime)\Icpr\nabla_{\o}^{\<i}\nabla_{\o}^{ j\>}\Phi(\chiprime)\,.
\end{eqnarray}

\paragraph{\bf \em Summary}

We have presented the distance-redshift relation to second order in general. Our only simplifying assumptions have been a flat background and equality of the first-order potentials (and zero primordial vectors and tensors). We have not assumed a matter model, and so our results hold for most dark energy models.

We have identified key new terms which govern gravitational lensing magnification for large over-densities~-- and under-densities where it has recently been shown that the linear lensing terms do not capture the full relativistic signal~\cite{Bolejko:2012uj}. In addition we have presented new effects which contribute to redshift space distortions. These are:
 
\paragraph{Nonlinear Doppler effect and transverse velocities} This comes in several forms. The radial parts of the scalar and vector second-order velocities contribute in the same way as at first order. Then the terms $\mathcal{O}(v^2,v\Phi)$ reveal the transverse Doppler contribution in the cosmological context:
\be
\Delta^v_\text{transverse}=\left(1- \frac{1}{\HH_s \chi_s}\right)\left[\nabla_\o ^iv_s+2\Ic\nabla_\o^i\Phi\right]\nabla_{\o i} v_s
\ee 
While small, these give the potential to measure transverse velocities through redshift space distortions and distance modifications. The prefactor in parentheses is large for low redshift amplifying the effect. Note that in the full Doppler contribution there is also a local dipole contribution, giving different lensing signals across the sky from our peculiar motion. 

\paragraph{Redshift space distortion and lensing coupling to Doppler lensing} Linear RSD and first-order gravitational lensing terms couple non-linearly to SW and ISW terms, but also to Doppler lensing which could be significant. In fact, the Doppler lensing contribution~\cite{Bonvin:2008ni,Bacon:2014uja}, including the main RSD and lensing corrections becomes (ignoring the second-order velocity which adds linearly to this)
\be
\kappa_v\approx -\left[\left(1- \frac{1}{ \chi_s\HH_s}\right)\left(1-\chi_s
\D_{\pp}^2v_s \right)
+\left(2-\frac{1}{\chi_s\HH_s}\right)\Ic\chi\nabla_{\o}^2\Phi
-\left(3-\frac{2}{\chi_s\HH_s}\right)\Ic\frac{(\chi-\chi_s)\chi}{\chi_s}\nabla^2_{\o}\Phi\right]\D_{\pp}v_s\,.
\ee
The new RSD contribution will alter and amplify the Doppler lensing at low redshift. 
The last term is the standard gravitational lensing term with integral kernel which peaks midway between source and observer, so will alter Doppler lensing at moderate redhsifts. The other lensing-type term contains a kernel which grows towards the source, meaning that for distant sources this will amplify the Doppler lensing correction.

\paragraph{Nonlinear density coupling} In the integrated contribution from $\Delta_\text{int}^{\kappa-\kappa}+\Delta_\text{int}^{\alpha-\nabla\kappa}+\Delta_\text{int}^{\Sigma-\Sigma}$ we have a product of the first-order lensing term and gradients of the gravitational  potential, which gives the  dominant  contribution to second-order lensing. [Roughly speaking, four $\nabla$'s and two $\Phi$'s give terms $\mathcal{O}(\delta^2)$.] 
In $\Delta_\text{int}^{\alpha-\nabla\kappa}$ we see coupling of transverse derivatives of the density fluctuations to the transverse velocity integrated along the line of sight. These offer the potential to measure transverse velocities as these terms can be very significant. 
The contribution from  $\Delta_\text{int}^{\kappa-\kappa}$  is a coupling between all the density fluctuations along the line of sight, and when the density contrast is $\mathcal{O}(1)$ can easily be comparable to the main first-order lensing term. Note that similar terms appear in the integrated shear term~$\Delta_\text{int}^{\Sigma-\Sigma}$. 

We can estimate the magnitude of these terms as follows. We can evaluate terms for a single lens at $\chi_\text{lens}$ with $\nabla_\o^2\Phi\approx 3\Omega_mH_0^2(1+z)\delta/2$, assuming a top-hat profile for $\delta$, of radius $R$. Then, the first-order lensing convergence becomes (to leading order in $R$, assuming $z$ is constant over the scale of the lens)
\be
\kappa^{(1)}\approx\frac{1}{3}\Omega_mh^2\frac{\chi_s-\chi_\text{lens}}{\chi_s}(1+z_\text{lens})\left(\frac{\chi_\text{lens}}{1\,\text{Gpc}}\right)
\left(\frac{R}{1\,\text{Mpc}}\right)
\left(\frac{\delta}{1000}\right)
\ee
which can approach unity for large over-densities. Using~\eqref{sdkflbjvskdvbs}, we can estimate, for one lens, 
\be
\Delta^{\kappa-\kappa}_\text{int}\approx \frac{1}{9}\Omega_m h^2(1+z_\text{lens})\left(\frac{R}{1\,\text{Mpc}}\right)^2
\left(\frac{\delta}{10^6}\right) \kappa^{(1)}
\ee
which forms part of the full second-order lensing convergence $\kappa\two$ ($\Delta_\text{int}^{\alpha-\nabla\kappa}$ and $\Delta_\text{int}^{\Sigma-\Sigma}$ will be a similar size). This is typically quite small per lens, but scales linearly with the number of lenses along a line of sight, so will build up significantly at high redshift.
The two lensing terms in the modifications to the Doppler lensing term are more significant still:
\be
\kappa_v\approx \kappa_v^{(1)}\left\{1-\left[
{\frac { \left( \chi_s-3\chi_\text{lens}\right) \chi_s\mathcal{H}_s-\chi_s+2\,\chi_\text{lens}}{\left(\chi_s- \chi_\text{lens}
 \right)(1-\chi_s\mathcal{H}_s )}}
\right]\kappa^{(1)}
\right\}
\ee
where $\kappa^{(1)}_v=-(1-1/\chi_s\mathcal{H}_s)\nabla\p v$. The term multiplying $\kappa^{(1)}$ is order unity, implying a significant correction to linear Doppler lensing behind over-densities. 


While many of the terms in the distance-redshift expansion are generally small, they offer a rich variety of new general relativistic effects to be understood. In particular, there are many terms which contribute to an ISW-type of effect, but now it involves integrals over the scalar, vector and tensor potentials at second-order, as well as the first-order potential squared, together with
its radial and tangential derivatives. Furthermore, we have also identified several instances of double integrated SW terms in both the redshift and the distance-redshift relation. These may be important in further refining our understanding of dark energy.

A final important relativistic effect which can be probed with the formalism presented above relates to the interpretation of the background model itself. Measurements of distances of supernovae, for example, are fitted to the background model via an all-sky average of the distance-redshift relation. This includes the monopole of the second-order corrections presented here, which may include non-trivial corrections to the background $D_A(z)$ relation. The size of this may be estimated via an ensemble average of the monopole of the $D_A(z)$ relation which does not vanish at second-order. There has been considerable debate as to the size of these corrections~-- see~\cite{Clarkson:2011zq} for a review. A  related lightcone average contribution was estimated in~\cite{BenDayan:2012ct,Ben-Dayan:2014swa,Nugier:2013tca} and found to be small, though potentially significant. This will be considered further in upcoming work~\cite{us}.

\paragraph{\bf\em Acknowledgments}

We thank David Bacon, Camille Bonvin and Ruth Durrer for useful discussions and  Giovanni Marozzi and Gabriele Veneziano for comments. Most of the computations here were done with the help of the tensor algebra package xPert/xAct \cite{Brizuela:2008ra} with its extension for homogeneous cosmology xPand~\cite{Pitrou:2013hga}.
OU and RM are funded by the South African Square Kilometre Array
(SKA) Project, CC and RM are supported by the National Research Foundation
(South Africa), RM is supported by the Science \& Technology Facilities Council (UK) (grant no.
ST/H002774/1), and all authors are supported by a Royal Society
(UK)/ NRF (SA) exchange grant.

\end{document}